\documentstyle[aps,epsf,preprint,prb]{revtex}
\tightenlines 
\begin{document}

\title{Mechanistic approach to generalized technical analysis of 
share prices and
stock market indices}

\author{ \large \bf M. Ausloos$^1$ and K. Ivanova$^2$ }

\address{ $^1$ GRASP and SUPRAS, B5, Sart Tilman, B-4000 Li\`ege, Belgium\\$^2$
Pennsylvania State University, \\ University Park, PA 16802, USA \\ }

%\preprint

%\draft

\maketitle

\begin{abstract}

Classical technical analysis methods of stock evolution are recalled, i.e. the
notion of moving averages and momentum indicators. The moving averages lead to
define death and gold crosses, resistance and support lines. Momentum 
indicators
lead the price trend, thus give signals before the price trend turns over. The
classical technical analysis investment strategy is thereby sketched. Next, we
present a generalization of these tricks drawing on physical principles, i.e.
taking into account not only the price of a stock but also the volume of
transactions. The latter becomes a time dependent generalized mass. 
The notion of
pressure, acceleration and force are deduced. A generalized (kinetic) energy is
easily defined. It is understood that the momentum indicators take into account
the sign of the fluctuations, while the energy is geared toward the absolute
value of the fluctuations. They have different patterns which are checked by
searching for the crossing points of their respective moving averages. The case
of IBM evolution over 1990-2000 is used for illustrations.

\end{abstract}

\vskip 2cm

\noindent {\it Keywords:} Econophysics; Force; Momentum; Energy; Pressure;
Acceleration; Mass; Volume; Price; Technical Analysis; Market Indices; Stocks

\section{Introduction}

Technical indicators as {\it moving average} and $momentum$ are part of the
classical technical analysis and are much used in efforts to predict market
movements.\cite{Achelis,FamaBlume,James} One question is whether 
these techniques
provide adequate ways to read the trends, and later on allow for an investment
strategy development. It has been shown that moving average trading 
rules can be
utilized \cite{Brock92,Hudson96} for USA and UK markets. In both 
futures and spot
foreign currency markets significant profits can be earned along these
lines.\cite{SzakmaryMathur} Parisi and Vasquez recently used the moving average
technique on emerging markets \cite{ParisiVasquez} to show that buy signals
consistently generate higher returns than sell signals. Moreover, returns
following sell signals are shown to be negative, which is not easily 
explained by
any of the currently existing equilibrium models. Related studies by 
Gunasekarage
and Power \cite{GP} also showed that technical trading rules have predictive
ability in South Asian stock markets.

One surprise to a physicist is the neglect of the volume of transactions in the
classical way of predicting the evolution of a share price or a 
market index. Yet
there was a sort of thermodynamic relationship between share price 
and exchanged
volume of shares in estimating the value of some company from an investor point
of view.\cite{phd,gendubl} Such an addition to measure and if possible predict
the evolution of stocks is introduced here. Indeed we present a 
generalization of
the classical technical analysis concepts taking into account the volume of
transactions. The latter becomes a {\it time dependent} or 
generalized mass. The
notions of pressure, acceleration and force are deduced. In that spirit, a
generalized (kinetic) energy is easily defined. It is pointed out that the
momentum correlations take into account the sign of the fluctuations, while the
energy is geared toward the absolute value of the fluctuations. They have
different patterns which are checked by searching for the crossing points of
their respective moving averages. The evolution of IBM share price and volume
between Jan 01, 1990 and Dec 31, 2000 is used for illustrations.

\section{Technical Analysis} \subsection{Moving Average} Consider a time series
$x(t)$ given at N discrete times $t$. The series (or signal) moving average
$M_{\tau}(t)$ over a time interval $\tau$ is defined as

\begin{equation} M_{\tau}(t)=\frac{1}{\tau}\sum_{i=t}^{t+\tau-1} 
x(i-\tau) \qquad
t=\tau+1,\dots,N \end{equation} i.e. the average of $x$ over the last 
$\tau$ data
points. For simplicity we suppose that the ticking times are equally 
spaced. One
can easily show that if the signal $x(t)$ increases (decreases) with time,
$M_{\tau}(t)<x(t)$ ($M_{\tau}(t)>x(t)$). Thus, a moving average 
captures the past
trend of the signal over a given time interval $\tau$. The IBM daily closing
value price signal between Jan 01, 1990 and Dec 31, 2000 is shown in 
Fig. 1 (top
figure) together with the $\tau=50$~day moving average taken from
Yahoo.\cite{yahooibm} The bottom figure shows the daily $volume$ of 
transactions
given in millions.

The moving average notion has already been discussed.\cite{nvmama,nvmav2} It is
obvious that like any other statistical $mean$, a moving average $M_{\tau}(t)$,
depends on the number of data points taken into account. There can be as many
moving averages as $\tau$ intervals. The shorter the $\tau$ interval the closer
to a signal is the moving average. However, a too short moving average may give
false messages about the long time trend of the signal. In Fig. 2 two moving
averages of the IBM signal for $\tau$=5 days (i.e. 1 week) and 21~days (i.e. 1
month) are compared for illustration.

The intersections of the $price$ signal with a moving average can define
so-called lines of $support$ or $resistance$.\cite{Achelis} A support
(resistance) line occurs when the local minimum (maximum) of $x(t)$ bounces on
$M_{\tau}(t)$. The lines are supposed to indicate the price level at which
$\tau$-investors believe that prices will move higher or lower respectively. In
Fig. 2, IBM lines of resistance happen e.g. around June 1993 and 
lines of support
around mid Oct. 1993 for the $\tau$ = 5 days and 21~days cases respectively.
Support and resistance levels depend on $\tau$ and are based in principle on
investment horizon strategy, but are in fact containing much 
psychological fancy.

Other features of the moving average prone investor framework are the
intersections between $two$ moving averages $M_{\tau_1}$ and $M_{\tau_2}$. They
might occur or not at drastic changes in the trends of $x(t)$.\cite{nvmama}
Consider again the two moving averages of IBM price signal for $\tau_1=5$~days
and $\tau_2=21$~days (Fig. 2). If $x(t)$ increases for a long period of time
before decreasing rapidly, $M_{\tau_1}$ will cross $M_{\tau_2}$ from 
above. This
event is called a {\it death cross} in empirical finance.\cite{Achelis} In
contrast, when $M_{\tau_1}$ crosses $M_{\tau_2}$ from below, the crossing point
is a {\it gold cross}. They appear more drastic when the respective slopes have
different signs. In Fig.2 a death cross and a gold cross occur near 
March 93 and
Oct. 93 respectively. The density of such crossing points between two moving
averages as a function of the difference in the characteristic 
$\tau$'s defining
the moving averages has been discussed elsewhere\cite{nvmama} and is shown in
Fig.3 for the case of interest here. Based on this idea, a new and efficient
approach has been suggested in order to estimate an exponent that characterizes
the roughness of a signal. From Fig.3 and Ref.\onlinecite{nvmama} the IBM
roughness exponent has been found equal to $0.44\pm0.02$ for the time interval
considered.

\subsection{Momentum Indicator}

The so called momentum\cite{Achelis} is another instrument of the technical
analysts. We will refer to it here as the {\it classical momentum} 
for reasons to
become obvious later. The classical momentum of a stock is defined over a time
interval $\tau$ as

\begin{equation} R_{\tau}(t)=\frac{x(t)-x(t-\tau)}{\tau}= 
\frac{\Delta x}{\Delta
t} \qquad t=\tau+1,\dots,N \end{equation}

For $\Delta t$ =$\tau$ = 1, the momentum is nothing else than the 
volatility. The
momentum $R_{\tau}$ for three time intervals, $\tau$=5, 21 and 
250~days, i.e. one
week, one month and one year, are shown in Fig. 4 for IBM between 
1990 and 2000,
together with a blow up for the years 1999-2000. The longer the period, the
smoother the momentum signal. Relevant information on the price trend turnovers
is usually considered to be found in a few {\it moving averages} of the
$momentum$, or {\it momentum indicator}, i.e. in

\begin{equation} R^{\Sigma}_{\tau}(t) = \sum_{i=t}^{t+\tau-1}
\frac{x(i)-x(i-\tau)}{\tau} \qquad t=\tau+1,\dots,N \end{equation} 
like those for
$\tau$ = 1 week, 1 month and 1 year. Notice that those are calculated over the
same time intervals over which the momentum is calculated. For IBM, these
$R^{\Sigma}_{\tau}$ are shown in Fig. 5. The density of intersections between
moving averages of these momenta could be calculated. The result is 
displayed in
Fig.3.

\subsection{Classical Strategy}

In Fig. 6 the IBM signal and its weekly (short-term), monthly (medium-term) and
yearly (long-term) moving averages are compared to the weekly (short-term),
monthly (medium-term) and yearly (long-term) momentum indicators to 
show the 1999
bullish and also beginning of bearish trends. The strategic message that is
coming out of reading the combination of these six indicators (Fig. 6) is that
one could start buying at some momentum bottom and sell at a maximum. A buy
position  is found for both monthly and weekly momentum indicators around (1)
Feb. 99. Another buy position occurs around mid-April 99 when the price surge
confirms the momentum uptrend, at a gold cross (2). Selling signals 
are given at
the maximum of the monthly momentum indicator in mid-Jan. 99 (6), 
near the second
half of mid-May (7) and mid-July (8). A death cross (3) occurs 
between the short
and medium term moving averages in Sept. 99 and is subsequently followed by the
maximum (4) of the monthly momentum, even turning negative. In Oct. 99, occurs
the maximum of the long-term momentum, just before the Oct. 99 crash 
(5).  A good
technical analyst would have  strongly recommended to sell the 
position since the
price is also falling down below the weekly moving average. Hope (or 
faith?) for
prospect reoccurs after Nov. 99 during one month.

\section{Generalized technical analysis}

Stock markets do have another component beside prices or volatilities. This is
the volume of transactions. It is introduced here as the {\it physical mass} of
stocks. Remember that the number of shares is constant over rather long time
intervals, i.e. usually between splits, like the mass of an object.

Consider $V(t)$ to be the volume of transactions of a stock with 
price $x(t)$ at
time $t$, - Fig. 1 (bottom). A generalized momentum $\widetilde 
R_{\tau}$ over a
time interval $\tau$ can be defined (see Fig. 7) as in physics through

\begin{equation} \widetilde R_{\tau}(t)=\frac{V(t)}{<V(t)>_{\tau}\tau}\cdot
\frac{x(t)-x(t-\tau)}{\tau}=m(t)\frac{\Delta x}{\Delta t}, \qquad
t=\tau+1,\dots,N \end{equation} where the total volume of transactions over the
interval $\tau$ is $<V>_{\tau}\tau=\sum_{i=1}^{\tau}V(i)$. In so doing, we
introduce some financial analogy to a {\it generalized time dependent mass}
$m(t)$ of a diffusing object. The total volume in the denominator is introduced
for a normalization purpose. Notice that we could choose other 
normalizations or
definitions of the mass or the generalized momentum : one could take the
$log(V(t))$, but this introduces a nonlinear transformation. One could use
$log[V(t+\tau)/V(t)]$ but that can be negative. One could normalize 
with respect
to $<V>_{\tau}$ only, but $\widetilde R_{\tau}(t)$ values would be larger, ....

In order to search for more definite indications on the stock trend changes,
represented as the complex influence between price and volume of 
transactions, we
further consider a {\it moving average of the generalized momentum}, 
i.e. a {\it
generalized momentum indicator} (Fig.8) which is

\begin{equation} \widetilde R^{\Sigma}_{\tau}(t)=\sum_{i=t}^{t+\tau-1}
\frac{V(t)}{<V>_{\tau}\tau} \cdot \frac{x(i)-x(i-\tau)}{\tau} \qquad
t=\tau+1,\dots,N \end{equation}

The IBM generalized momentum indicator $\widetilde R_{\tau}$ for three time
intervals $\tau$, i.e. 1 week, 1 month and 1 year and their 
corresponding moving
average $\widetilde R^{\Sigma}_{\tau}$ are shown in Fig. 8. It is observed that
the multiplicative factor results only in enhancing features, i.e. 
the mass being
here positive. However the slopes in the generalized momentum 
indicator are much
enhanced due to volume variations. A comparison of moving averages of the IBM
signal and its IBM generalized momentum for three different horizons together
with the volume of transactions are shown in Fig. 9. The volume of transactions
is plotted in tenths of millions on the generalized momentum base line. The
long-term (yearly) generalized momentum signal is marked by dot-dash curve and
has small positive values, with a steady increase up to the maximum of the IBM
price around mid Sept. 1999 marked by the index (3). See the well 
marked crash of
Oct. 99; also displayed in Fig.9, just after (5). The crash coincides with the
maximum in the volume of transactions, and a negative momentum.

A closer look of the short (weekly) and medium-term (monthly) momentum in terms
of sell/buy message point of views again suggests to buy around Feb. 1999 (see
Fig. 6 as well) at the minimum of the momentum curve. The new feature 
is that the
peaks of the generalized medium term (monthly) and long-term (yearly) momentum
enhances several changes in the price trend represented by its 
monthly or yearly
moving average. Peaks are observed at mid Jan. 99 (6), mid May (7) (after the
gold cross (2) ) and mid July 1999 (8). Notice that (3) and (8) 
coincide with the
monthly momentum turning to negative values after reaching a weak 
maximum. Notice
the differences between Fig. 6 and Fig. 9, i.e. the enhanced, and even modified
structures, between (6) and (1),  and between (4) and the Oct. crash, 
due to the
volume effect.

\subsection{Pressure, Acceleration and Force Indicators}

Introducing a ''mass'' $m(t)=V(t)/(<V>_{\tau}\tau)$ and the ''velocity''
$v(t)=\Delta x/\tau$ through the standard notion of physical momentum

\begin{equation} \widetilde 
R^{\Sigma}_{\tau}=p(t)=\sum_{i=t}^{t+\tau-1} m(i)v(i)
\end{equation} enables us to take into account both sides of the {\it market
coin} and their impact on one another.

Note that $<V>_{\tau}\tau$ ensures to include a $pressure$\footnote{ A global
kinetic theory for prices has been derived considering an equilibrium market
(with actors having all identical relaxation times). In closing the set of
equations, an equation of state, with a pressure and a temperature, between the
price, as the order parameter of a stock, and the volume of exchanged 
shares were
introduced.\cite{maboltzmann}} contribution from the variation of the 
accumulated
volume of transactions during the time period $\tau$ on the change of the
momentum of the stock. From a mathematical point of view at the time 
of the peaks
of generalized momentum curve $\widetilde R^{\Sigma}_{\tau}$, the first
derivative of the momentum are obviously equal to zero and the second 
derivative
is negative or positive depending whether there is a peak or a dip. Because of
the time dependent ''mass'' this implies that

\begin{equation} p'(t)=\sum_{i=t}^{t+\tau-1} m'(i)v(i) + \sum_{i=t}^{t+\tau-1}
m(i)v'(i)=0 \end{equation}

or

\begin{equation} p'(t)=\sum_{i=t}^{t+\tau-1} m'(i)v(i) + \sum_{i=t}^{t+\tau-1}
m(i)a(i)=0, \label{eqp} \end{equation} where $a(i)$ can be thought as an
$acceleration$.\footnote{Price velocity and price acceleration are two
fundamental indicators which of course already exist in the economy literature
and econophysics, e.g. they were recently used to construct a general
classification of market indices possible patterns deviating from the random
walk.\cite{sornettetechanal}} Therefore the second term in Eq. \ref{eqp}
represents a classical $force$ that acts upon the object (stock or 
market index).
It originates from the speed of change of stock price whose sign is either
positive or negative. At times of extrema this force balances the first term in
Eq. \ref{eqp}, determining the rate in the volume of transactions. The sign of
this first term depends on the accumulated sum of $m'(i)v(i)$, since 
both $m'(i)$
and $v(i)$ can be negative or positive.

Therefore, the fact that the first derivative of the momentum $p'(t)$ 
vanishes at
time $t_1$ can lead to a deeper reading of the price/volume interactions. The
maximum of the (generalized or not) momentum indicator at $t_1$, mid Jan. 1999,
mark (6) in Fig. 6 and Fig.9, is related to a sharp increase of the volume of
transactions, as the rescaled volume of transactions is of the order of
$\widetilde R^{\Sigma}_{\tau}$. For Eq. \ref{eqp} to be satisfied, the second
term in the equation should be negative. Thus the derivative of the $velocity$,
i.e. the acceleration $\Delta v(t)/\Delta t$ should be negative, which means a
change in the derivative of the price trend. This is easily seen at the
resistance point, that also coincides with a death cross between the IBM signal
and its moving averages for the one week and one month time averaging 
at (1). The
same discussion pertains to the maximum of the momentum curve in mid July 1999,
mark (8) in Fig. 6 and Fig. 9, when the IBM share price breaks the resistance
line, just before a death cross when the price drops below its monthly moving
average, while the weekly moving average becomes negative.

However, the cause of the peak in the generalized momentum indicator 
at mid-May,
1999, mark (7) in Fig. 6 and Fig. 9, is more complex than for mid-Jan. 99, mark
(6), or mid-July 99, mark (8). Indeed the acceleration during most of the two
prior months is positive and so is the second term in Eq. \ref{eqp}. 
At the time
of the price maximum on May 13, 1999 the classical force has to be (and is)
balanced by a (negative) contribution from the first term. Because the velocity
$v(i)$ is positive during that period, the derivative of the volume of
transactions (the mass) has to be (and is) most of the time negative.

Thus the increase of momentum is not entirely due to the price 
increase, but does
have another component, hidden if one does not consider the above 
generalization.
After May 13, 1999, the price drops to the monthly moving average and 
rebounds on
a support line. Thus this can be interpreted as a continuous price 
increase with
corrections to the moving average price due to the influence of the volume of
transactions. The observed evolution of the generalized momentum of 
the IBM stock
implies that some {\it generalized force} can be considered as a cause of these
changes.

\subsection{Energy}

Mechanically speaking it can be thought that some $energy$ is also accumulated
through the interplay between the price and the volume of 
transactions and causes
a generalized force to act. Therefore, a kinetic ''energy'' can be 
introduced as

\begin{equation} E^{\Sigma}_{\tau}(t) = \sum_{i=t}^{t+\tau-1}
\left(\frac{x(i)-x(i-\tau)}{\tau}\right)^2 \qquad t=\tau+1,\dots,N.
\end{equation}

The generalized mass introduced above leads also to generalize the 
''energy'' of
the stock signal to be like in mechanics (we drop the factor $1/2$)

\begin{equation} \widetilde E^{\Sigma}_{\tau}(t) =
\frac{1}{\tau}\sum_{i=t}^{t+\tau-1} \frac{V(t)}{<V>_{\tau}} \cdot
\left(\frac{x(i)-x(i-\tau)}{\tau}\right)^2 \qquad t=\tau+1,\dots,N 
\end{equation}
or \begin{equation} \widetilde E^{\Sigma}_{\tau}(t) = m(t)\left(\frac{\Delta
x}{\Delta t}\right)^2=mv^2. \end{equation}

Theoretical and generalized $energy$ of the IBM stock are shown in Figs. 10-11
for the same three usual time interval averages $\tau$, i.e. one 
week, one month,
one year. Note (insert of Fig. 10) two large peaks in the monthly theoretical
energy : the high one in mid-May 1999 corresponding to (7) in Fig. 9, 
and one at
the crash time in October. A small peak in mid-July, mark (8) in Fig. 9,
corresponds to the maximum of the price and the generalized momentum. The
appearance of a huge increase in energy in Oct. 99 together with the negative
slope of the generalized momentum indicator should be interpreted now as a
serious warning for an incoming crash. A mimimum in kinetic energy 
can be seen to
provide information on strong continuing price increase or decrease depending on
the momentum rate of change. The local maximum of a kinetic 
energy as in
classical physics indicates some accumulation of energy at that time for the
evolution of the stock, - accumulation which must be dissipated ! It is
interesting to note that the monthly generalized energy (Fig. 11, inset)
emphasizes even further the huge accumulation of energy in Oct. 1999.

\section{Conclusions}

A thermodynamic-like phase diagram between the daily closing price 
($P$) and the
daily transaction volume ($V$) resulted in a fundamental phase diagram for
companies quoted on stock exchanges.\cite{phd} This pointed out to search for
correlations between volume and price. In the same spirit of econophysics
connecting physics and financial data for share prices or stock 
market indices we
have developed a generalization of the classical method by technical analysts,
i.e starting from the notion of moving averages and momentum
indicators.\cite{Achelis} One can take into account not only the 
price of a stock
but also the volume of transactions which is similar to a time dependent
generalized mass in mechanics. This might interestingly serve for generalizing
ideas on anomalous diffusion model(s)\cite{plerou} for price 
evolutions as well.
A notion of pressure, acceleration and force concepts are deduced and 
identified
to their classical mechanics counterparts. The terms have sometimes 
been used in
the literature but outside their classical meaning.\cite{castiglione} A
generalized (kinetic) energy is also easily defined, in the same lines of an
analogy. A correspondence between mechanical terms, classical 
technical analysis
terms and to be seen generalized concepts of technical analysis is 
found in Table
I. This might help in devising a free energy like approach, as also introduced
elsewhere.\cite{AJP}

Introducing the product of price and mass of transactions in indicators renders
the new signals very rough. The number of intersections, mimima and maxima is
highly modified.   Whether the above generalization of technical analysis
concepts can be used to increase the number of buy/sell orders, and develop new
strategies through nine rather than six indicators (shown as a summary in Figs.
12-13) is left for further work. The interest of such considerations will be
obvious if cumulated (good !) predictions\cite{cumulatedpredict} from the above
stand over usual findings.

Finally it seems that the (generalized) momentum and energy concepts so
introduced concern other effects studied in economy, i.e. the correlation
existence of signs and amplitudes of share price variations. It is easily
understood that the momentum (and momentum correlations) take into account the
sign of the fluctuations (and their correlations), while the energy (and energy
correlations) is geared toward the absolute value of the fluctuations 
(and their
correlations). In thermodynamics of non-equilibrium processes, these 
correlations
are described by the $viscosity$ and {\it thermal conductivity} coefficient
respectively. Thus the above concepts might also serve in a dynamic equation
framework.

\vskip 2.0cm {\bf Acknowledgements}

MA thanks Etienne Labie and Denis Vanderborght (Leleux Ass.) for practical
comments and references.

\vskip 1cm

\begin{table}[ht] \begin{center} \begin{tabular}{|c|c||c|c||c|c||} \hline
classical & & & classical &&generalized \\ mechanics&& & technical analysis& &
technical analysis \\ \hline mass&$m(t)$&$V(t)$&(Volume)&$V(t)$&Volume\\ \hline
distance&$x(t)$&$x(t)$&price&$x(t)$&price\\ \hline trend&$<x(t)>$&$<x(t)>$
&moving average&$<x(t)>$& moving average\\ \hline momentum&$m(t)v(t)$& $\Delta
x(t)/\Delta t$& momentum&$V(t)x'(t)$&generalized \\ &&&&&momentum\\ \hline
kinetic &$m(t)v^2(t)$&$<(\Delta x(t)/\Delta
t)^2>$&(kinetic&$V(t)x'^2(t)$&generalized \\ energy&&&energy)&&kinetic energy\\
\hline \end{tabular} \end{center} \end{table}

\vskip 1cm 
\newpage \begin{figure}[ht] \begin{center} \leavevmode \epsfysize=8cm
\epsffile{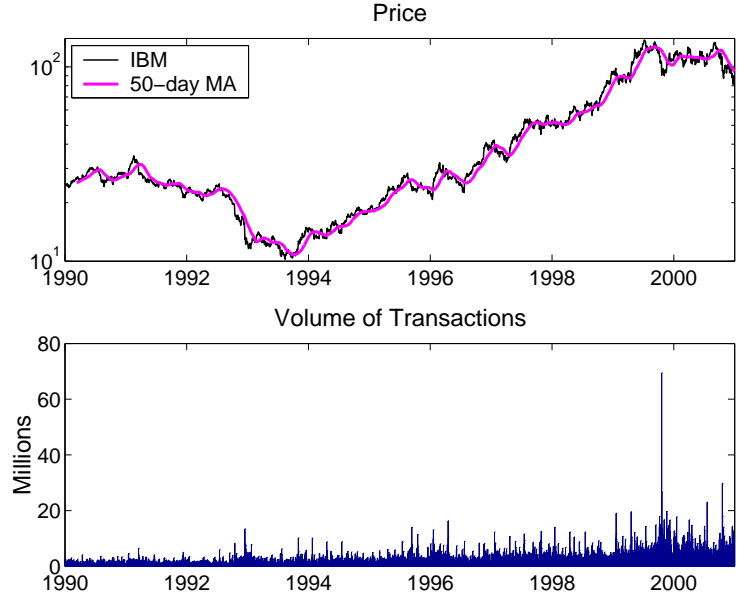} \end{center} \caption{(top) IBM daily closing value signal
between Jan. 01, 1990 and Dec. 31, 2000, i.e. 2780 data points with $yahoo$
moving averages for $\tau$ = 50 d; (bottom) daily volume of transactions in
millions\cite{yahooibm} } \end{figure}

\begin{figure}[ht] \begin{center} \leavevmode \epsfysize=8cm 
\epsffile{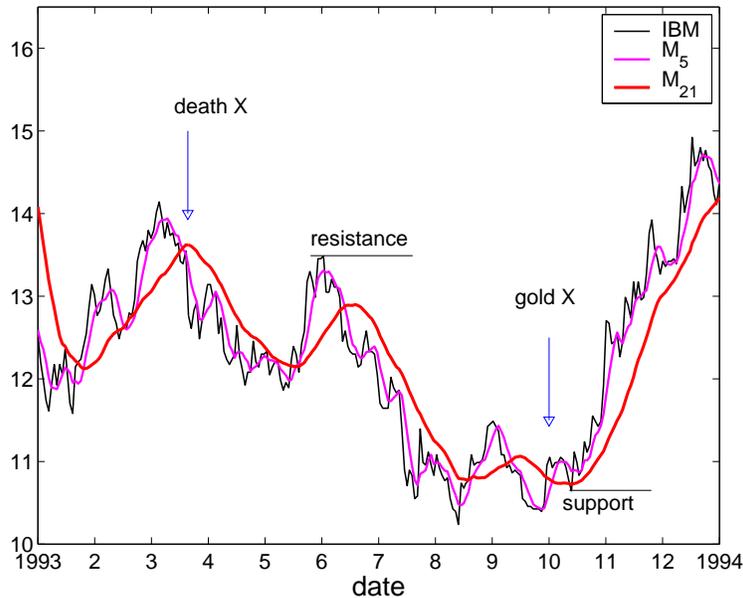}
\caption{Typical IBM daily closing value signal between Jan. 01, 1993 and Dec.
31, 1993, with two moving averages, $M_{\tau_1}$ and $M_{\tau_2}$ for
$\tau_1=5$~days and $\tau_2=21$~days. Death and gold crosses, resistance and
support lines are defined in the text } \end{center} \end{figure}

\newpage \begin{figure}[ht] \begin{center} \leavevmode \epsfysize=8cm
\epsffile{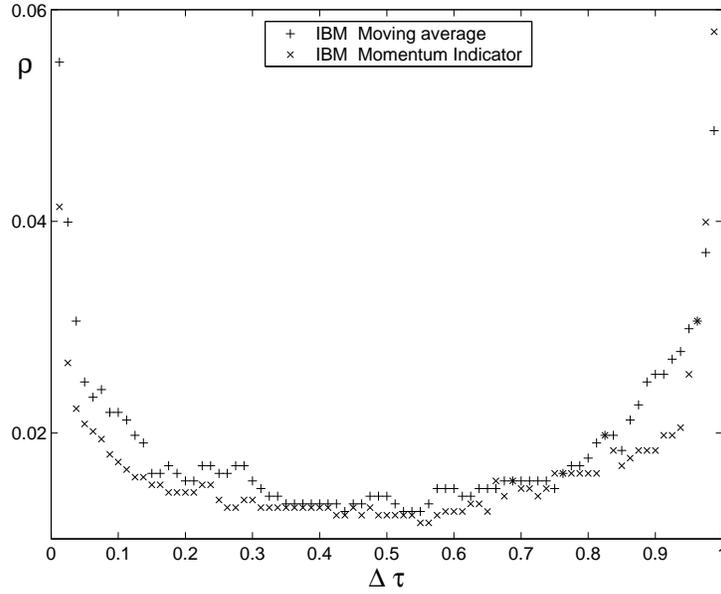} \caption{pluses: density $\rho$ of crossing points between
two moving averages with $\Delta \tau=(\tau_1-\tau_2)/\tau_2$ and fixed
$\tau_2=80$ from IBM daily closing value signal between Jan 01, 1990 
and Dec 31,
2000; crosses : same for momentum indicators} \end{center} \end{figure}

\begin{figure}[ht] \begin{center} \leavevmode \epsfysize=8cm 
\epsffile{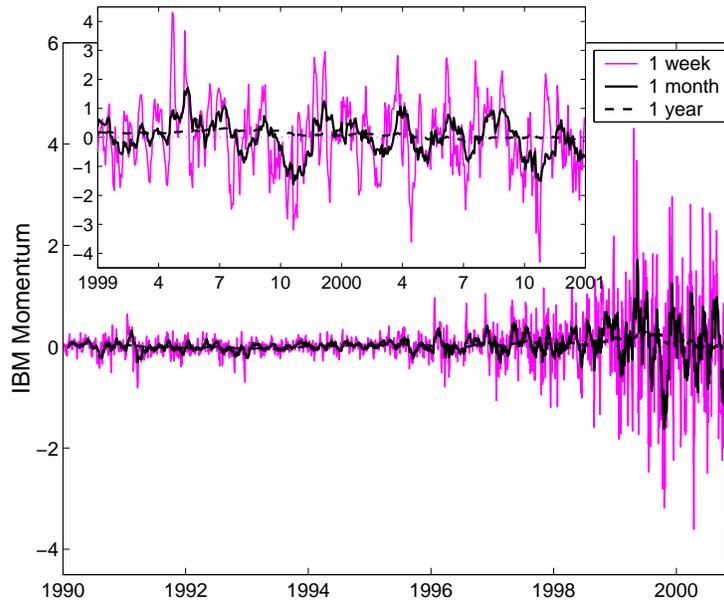}
\caption{ IBM 1 week, 1 month and 1 year $physical$ momentum between Jan. 01,
1990 and Dec. 31, 2000. Insert : blow up of the last two year time interval }
\end{center} \end{figure}

\newpage \begin{figure}[ht] \begin{center} \leavevmode \epsfysize=8cm
\epsffile{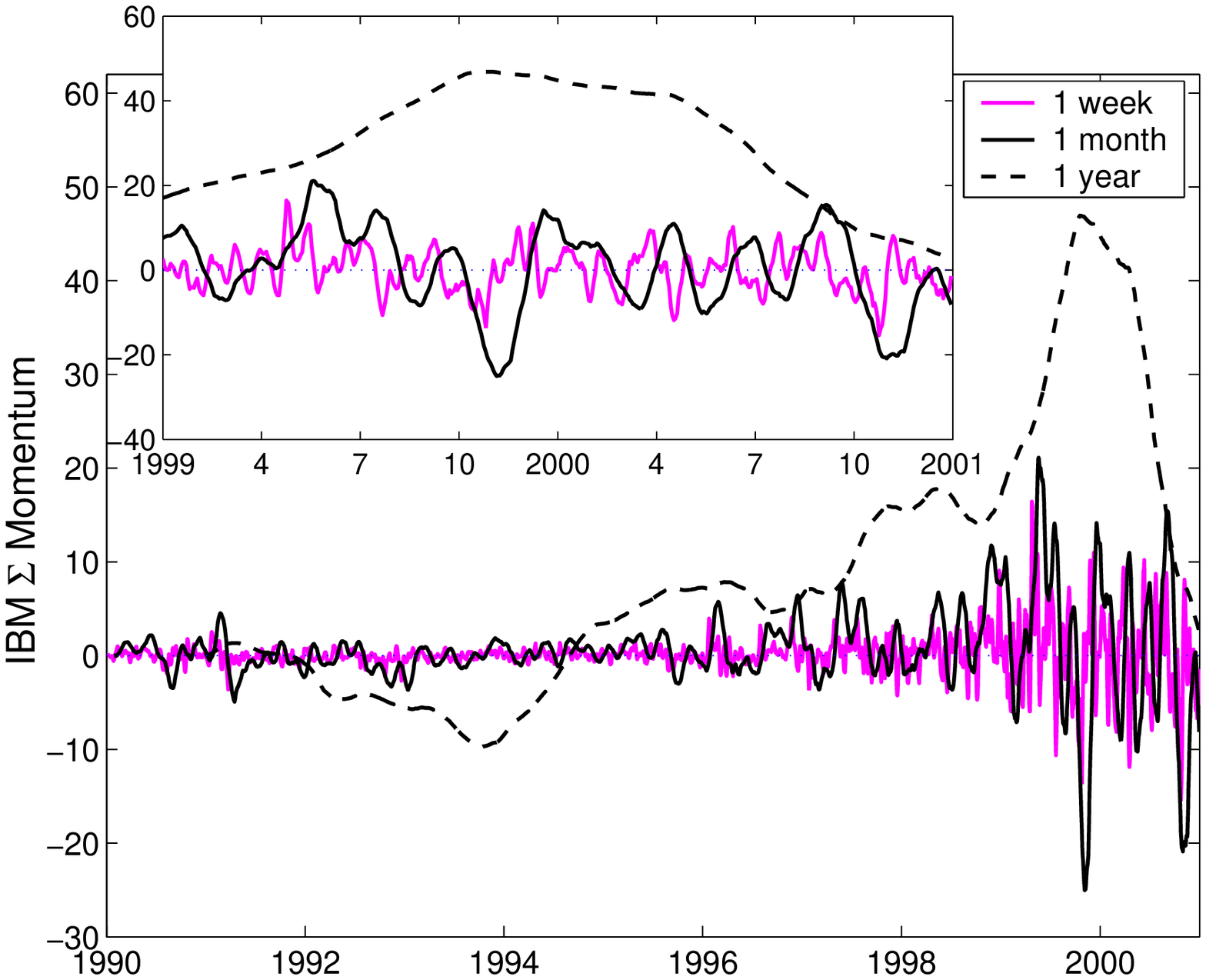} \caption{IBM momentum indicators for three different time
periods $\tau$, 1 week (light gray curve), 1 month (gray curve) and 1 
year (dash
curve). Insert : blow up of the last two year time interval} \end{center}
\end{figure}

\begin{figure}[ht] \begin{center} \leavevmode \epsfysize=8cm 
\epsffile{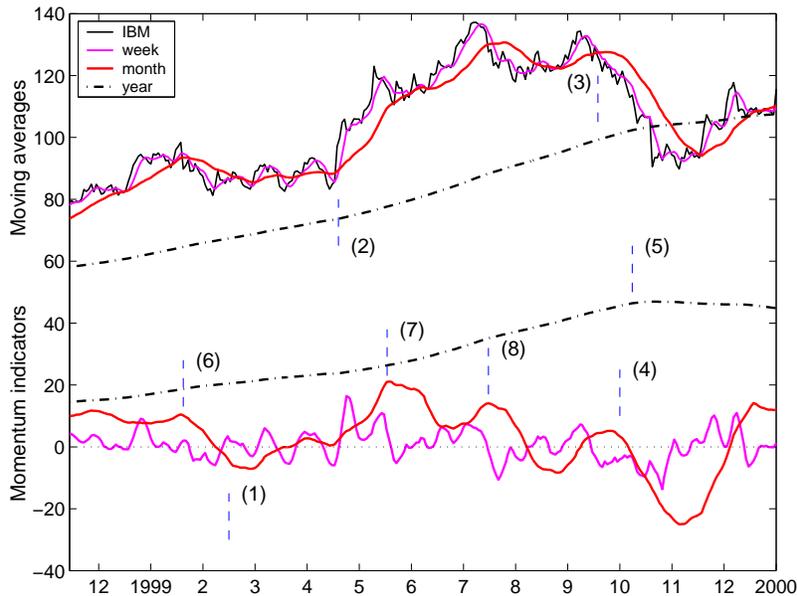}
\caption{IBM signal, moving averages of IBM signal and its classical momentum
indicators in 1999 or so for three time horizons, short-term (weekly) 
(light gray
curve), medium-term (monthly) (gray curve) and log-term (yearly) (dot-dash
curve), for discussing classical investment strategy; for lines at (1)-(5) see
text} \end{center} \end{figure}

\newpage \begin{figure}[ht] \begin{center} \leavevmode \epsfysize=8cm
\epsffile{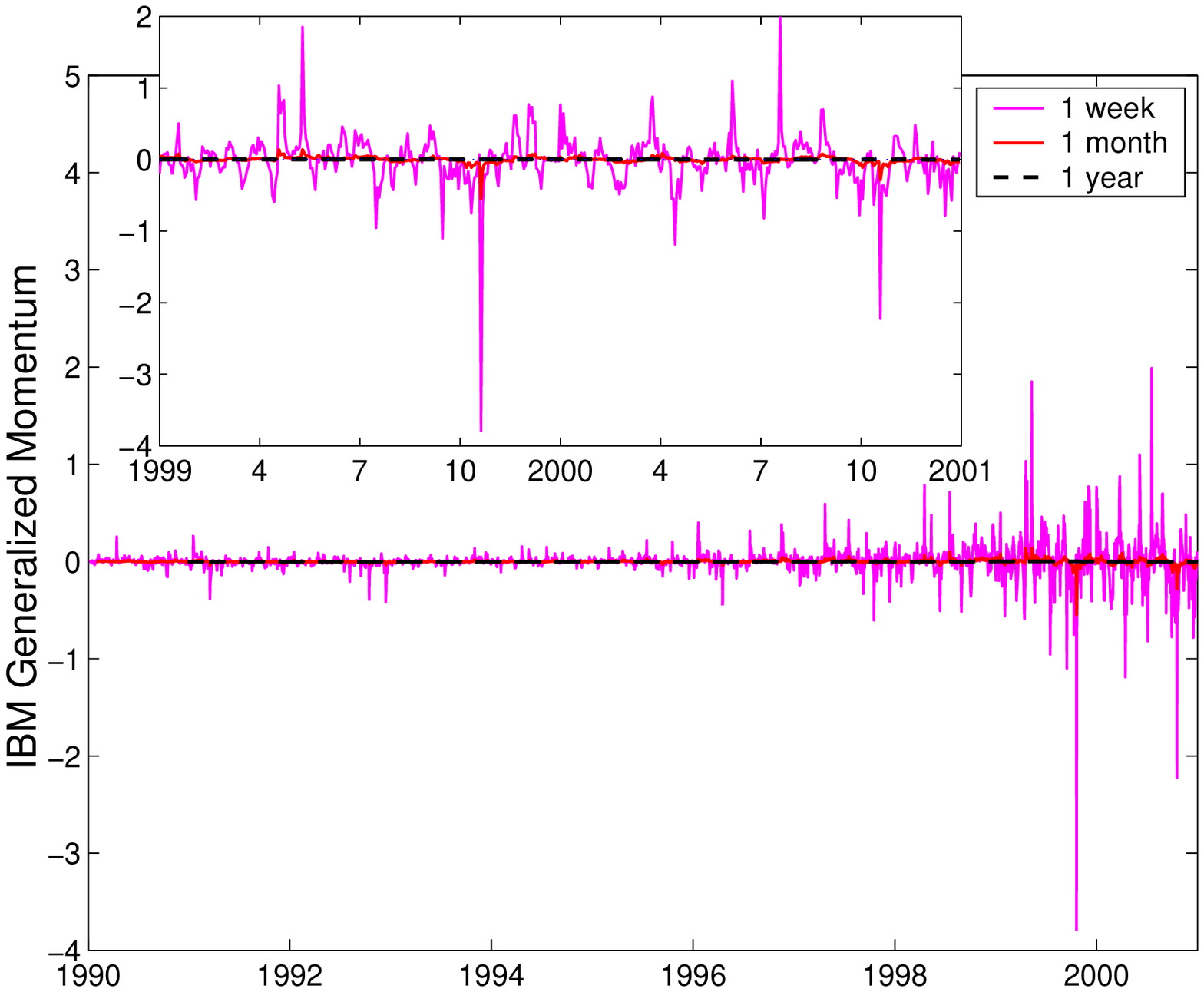} \caption{IBM generalized momentum between Jan 01, 1990 and
Dec 31, 2000 for three different time periods $\tau$, 1 week (light 
gray curve),
1 month (gray curve) and 1 year (dash curve). Insert : blow up of the last two
year time interval} \end{center} \end{figure}

\begin{figure}[ht] \begin{center} \leavevmode \epsfysize=8cm 
\epsffile{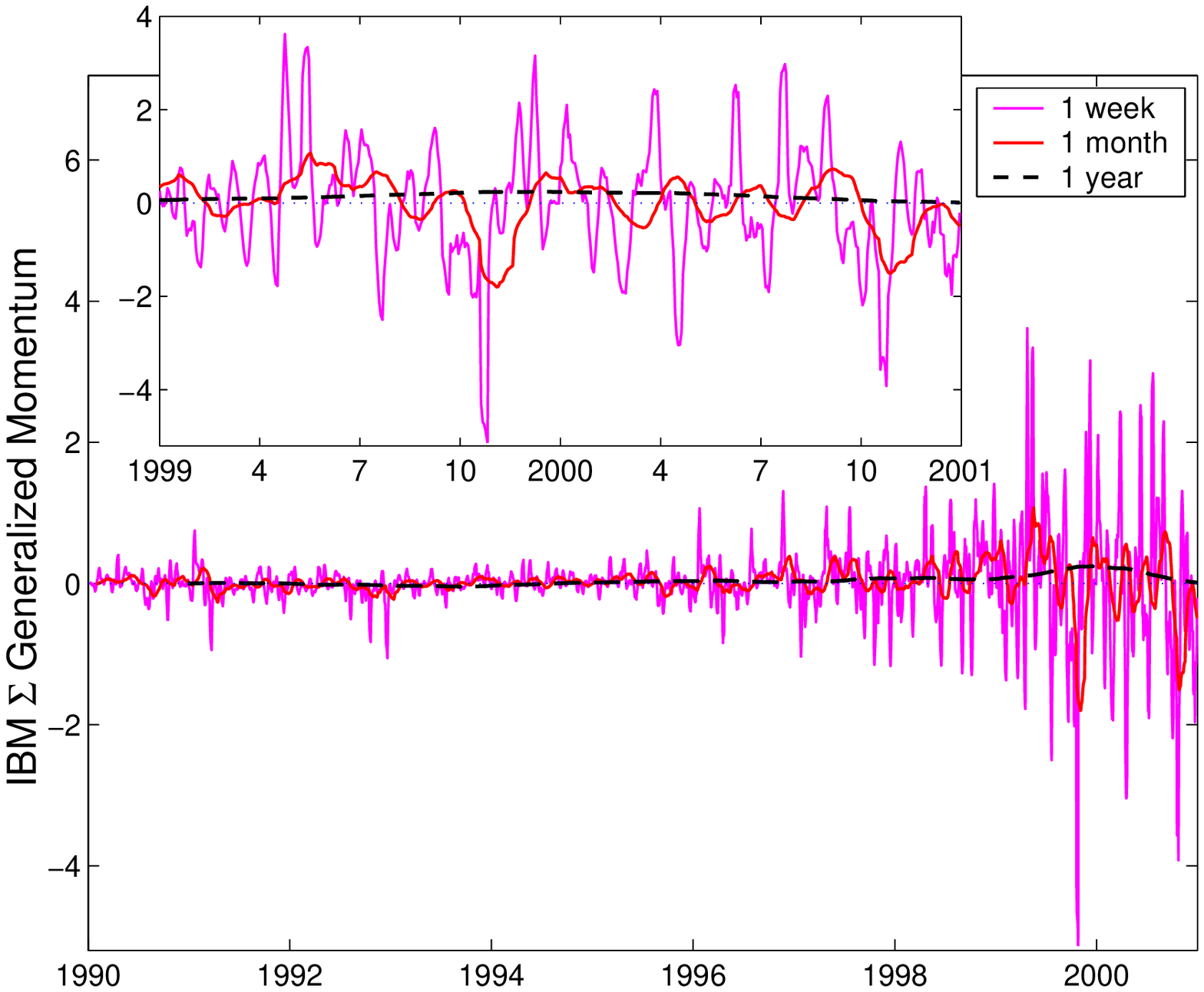}
\caption{IBM generalized momentum indicators between Jan 01, 1990 and Dec 31,
2000 for three different time periods $\tau$, 1 week (light gray 
curve), 1 month
(gray curve) and 1 year (dash curve). Insert : blow up of the last 
two year time
interval} \end{center} \end{figure}

\newpage \begin{figure}[ht] \begin{center} \leavevmode \epsfysize=8cm
\epsffile{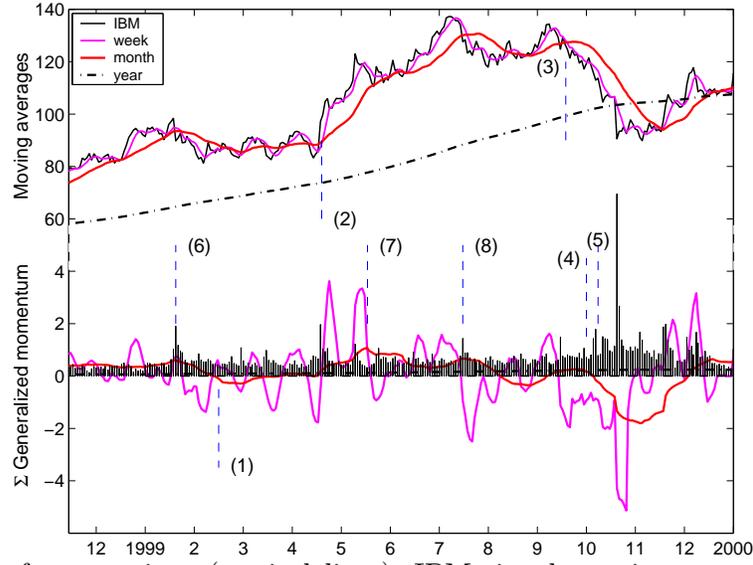} \caption{Volume of transactions (vertical lines), IBM
signal, moving averages of IBM signal and its generalized momentum 
indicators in
1999 for three time horizons, short-term (weekly) (light gray curve), 
medium-term
(monthly) (gray curve) and log-term (yearly) (dot-dash curve), for interpreting
price evolution} \end{center} \end{figure}

\begin{figure}[ht] \begin{center} \leavevmode \epsfysize=8cm
\epsffile{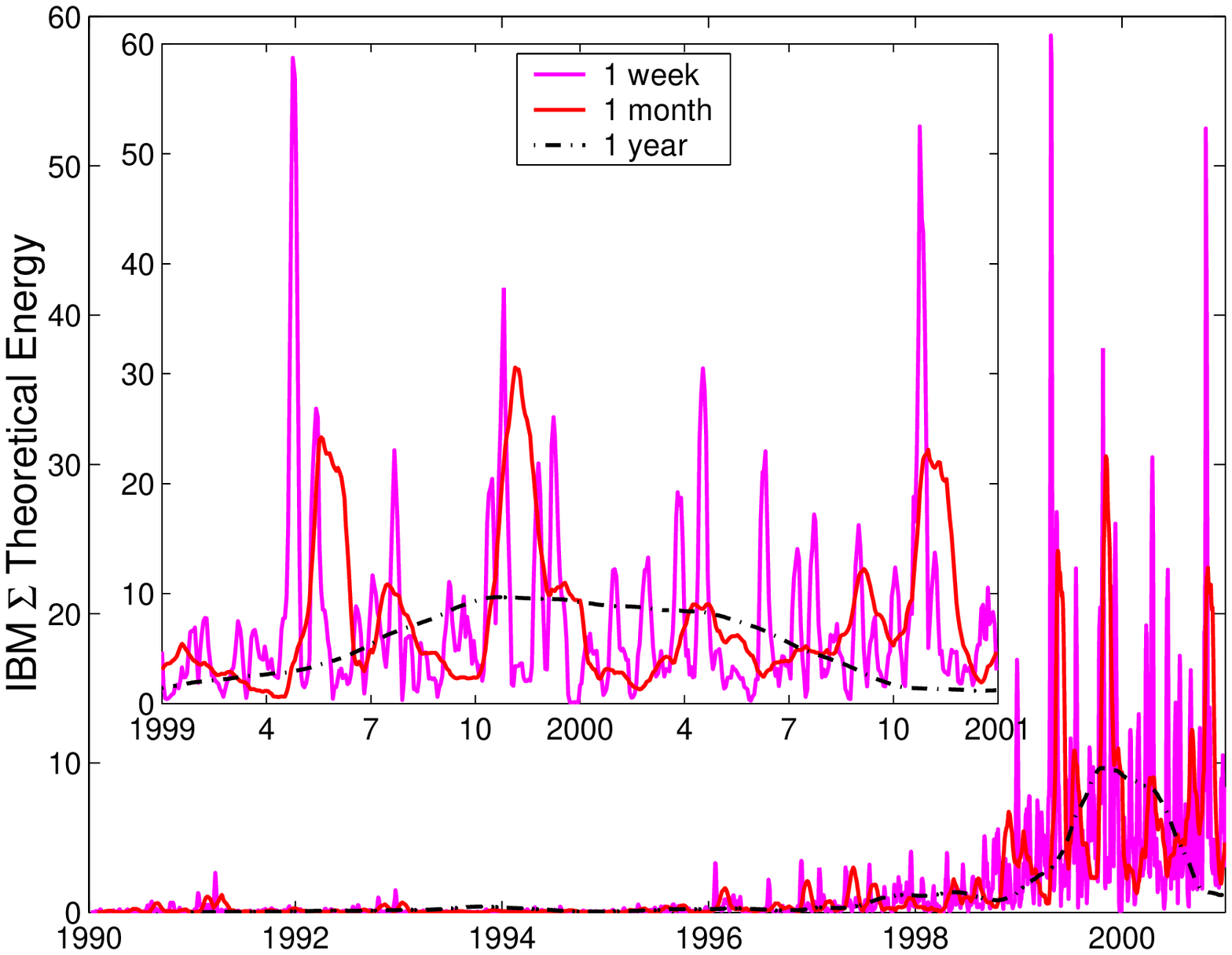} \caption{Theoretical (kinetic) energy of IBM between Jan
01, 1990 and Dec 31, 2000 for three time horizons, short-term (weekly) (light
gray curve), medium-term (monthly) (gray curve) and log-term (yearly) (dot-dash
curve); Insert : blow up of the last two year time interval} \end{center}
\end{figure}

\newpage \begin{figure}[ht] \begin{center} \leavevmode \epsfysize=8cm
\epsffile{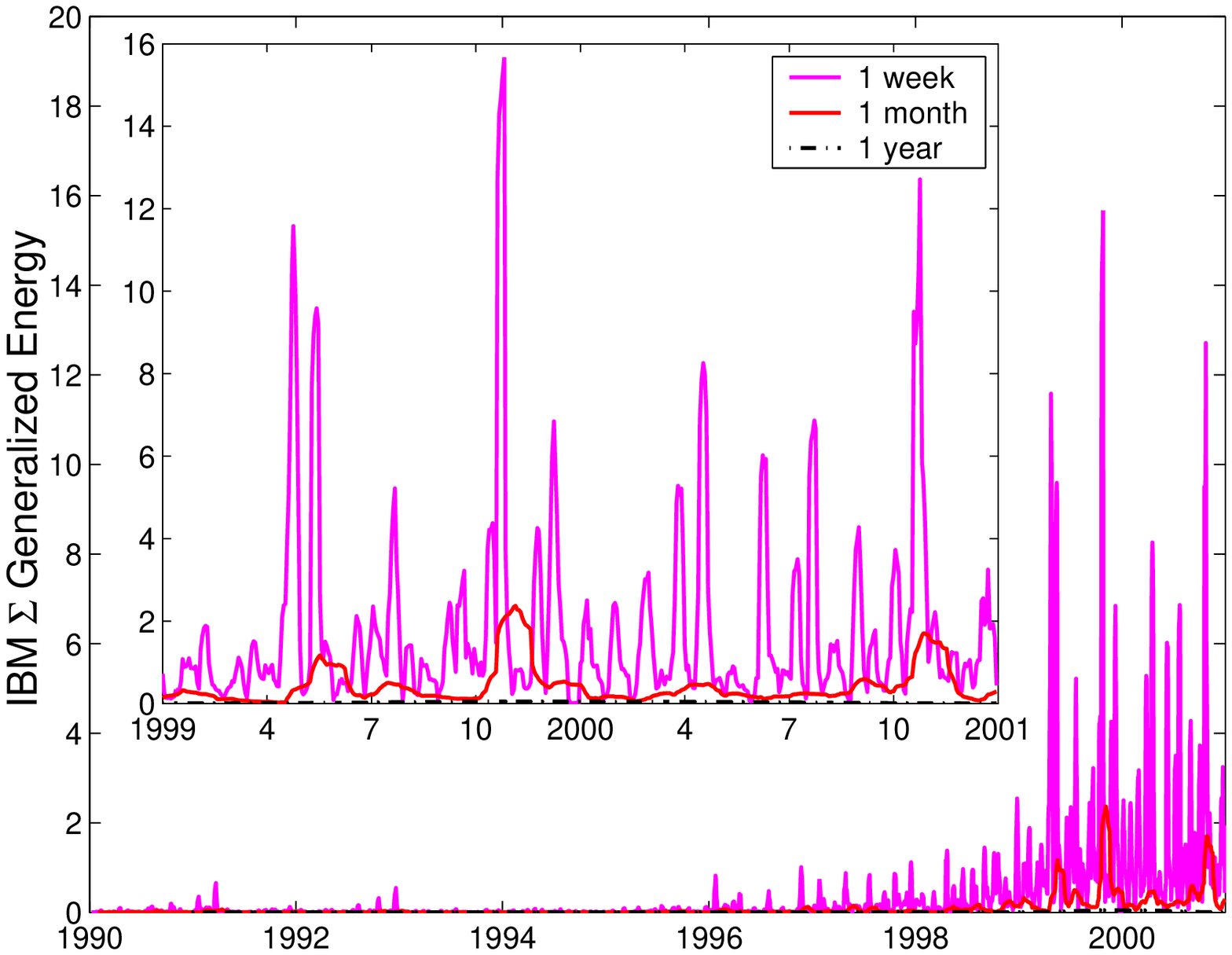} \caption{Generalized (kinetic) energy of IBM between Jan
01, 1990 and Dec 31, 2000 for three time horizons, short-term (weekly) (light
gray curve), medium-term (monthly) (gray curve) and log-term (yearly) (dot-dash
curve); Insert : blow up of the last two year time interval} \end{center}
\end{figure}

\begin{figure}[ht] \begin{center} \leavevmode \epsfysize=8cm
\epsffile{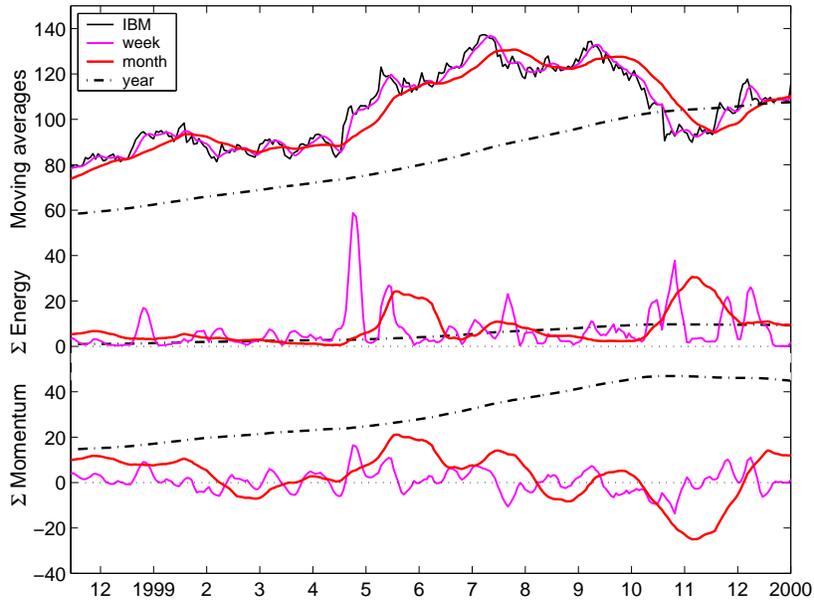} \caption{ Comparison of moving averages, (kinetic) energy
and momentum indicator of IBM in 1999 for three time horizons, short-term
(weekly) (light gray curve), medium-term (monthly) (gray curve) and log-term
(yearly) (dot-dash curve) in view of elaborating an investment strategy}
\end{center} \end{figure}

\newpage \begin{figure}[ht] \begin{center} \leavevmode \epsfysize=8cm
\epsffile{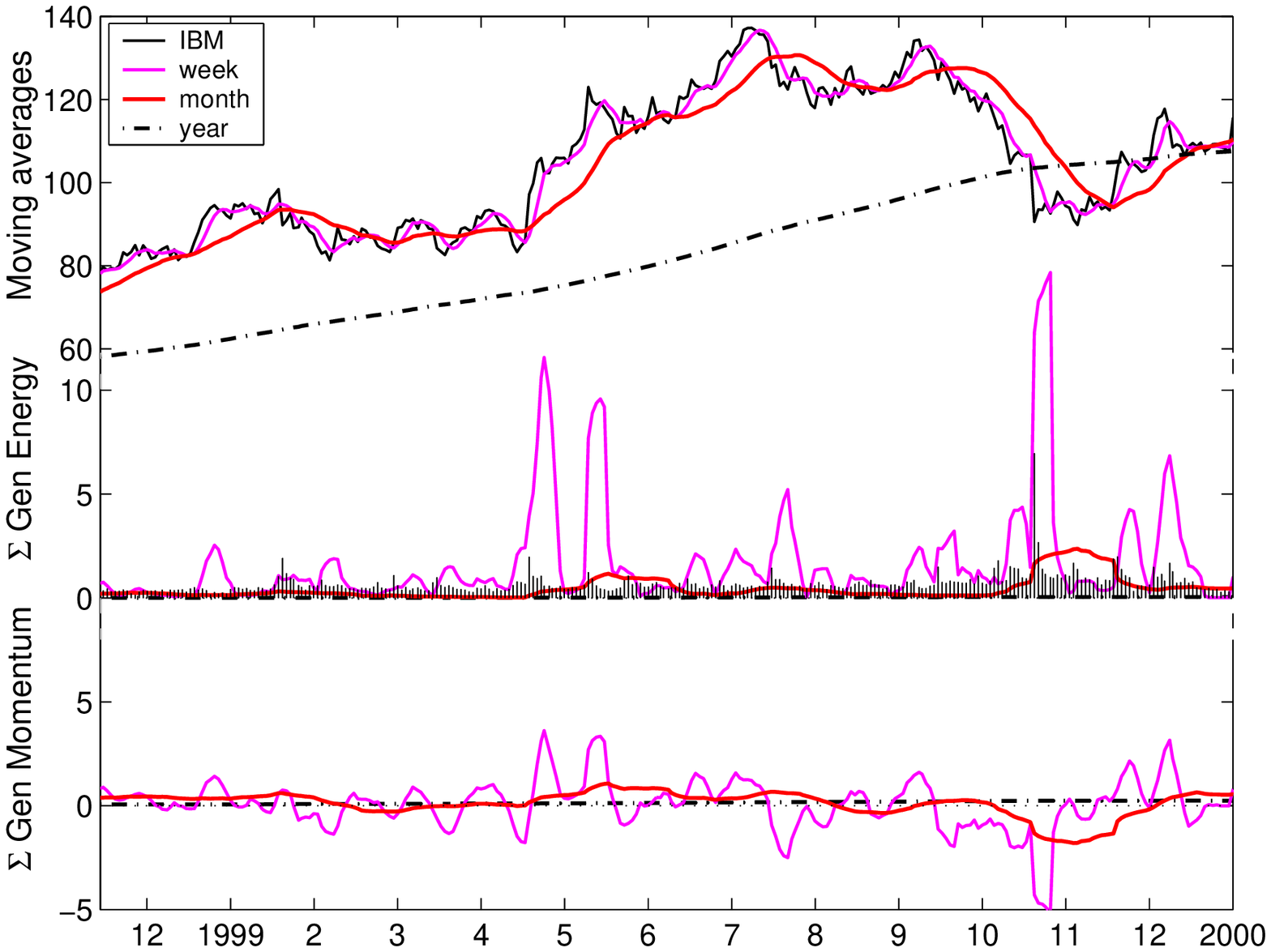} \caption{Comparison of moving averages, generalized
(kinetic) energy and generalized momentum indicator of IBM in 1999 
for three time
horizons, short-term (weekly) (light gray curve), medium-term (monthly) (gray
curve) and log-term (yearly) (dot-dash curve) in view of elaborating an
investment strategy; the volume of transaction is displayed on the same line as
the generalized energy} \end{center} \end{figure}


\begin{thebibliography}{99}

\bibitem{Achelis} see S.B. Achelis, in http://www.equis.com/free/taaz/

\bibitem{FamaBlume} E.F. Fama, M. Blume, {\it J. Bus.} {\bf 39}, 226 (1966)

\bibitem{James} F.E. James Jr., {\it J. Financ. Quant. Anal. } {\bf 3}, 315
(1968)

\bibitem{Brock92} W. Brock, J. Lakonishok, B. LeBaron, {\it J. 
Finance} {\bf 47},
1731 (1992)

\bibitem{Hudson96} R. Hudson, M. Dempsey, K. Keasey, {\it J. Banking Finance}
{\bf 20}, 1121 (1996)

\bibitem{SzakmaryMathur} A.C. Szakmary, I. Mathur, {\it J. Int. Money Finance}
{\bf 16}, 513 (1997)

\bibitem{ParisiVasquez} F. Parisi, A.Vasquez, {\it Emerging Markets 
Review} {\bf
1}, 152 (2000)

\bibitem{GP} A.Gunasekarage, D. M. Power, {\it Emerging Markets 
Review} {\bf 2},
17 (2001)

\bibitem{phd} K. Ivanova, {\it Physica A} {\bf 270}, 567 (1999)

\bibitem{gendubl} M. Ausloos, K. Ivanova, unpublished

\bibitem{yahooibm} http://finance.yahoo.com

\bibitem{nvmama} N. Vandewalle, M. Ausloos, Ph. Boveroux {\it Physica A} {\bf
269}, 170 (1999)

\bibitem{nvmav2} N. Vandewalle, M. Ausloos, {\it Phy. Rev E} {\bf 58}, 6832
(1998)

\bibitem{maboltzmann} M. Ausloos, {\it Physica A} {\bf 284}, 385 (2000)

\bibitem{sornettetechanal} J.V. Andersen, S. Gluzman, D. Sornette, {\it Eur.
Phys. J. B} {\bf 14}, 579 (2000)

\bibitem{plerou} P. Gopikrishnan, V. Plerou, X. Gabaix, L.A.N. Amaral, H.E.
Stanley, {\it Physica A} {\bf 299}, 137 (2001)

\bibitem{castiglione} F. Castiglione, R.B. Pandey, D. Stauffer,{\it Physica A}
{\bf 289}, 223 (2001)

\bibitem{AJP} W. M. Saslow, {\it Am. J. Phys. } {\bf 67}, 1239 (1999)

\bibitem{cumulatedpredict} J.G. de Gooijer, A. Klein, {\it Intern. J. 
Forecast. }
{\bf 7}, 501 (1992)




\end{thebibliography}
\end{document}